\documentstyle[preprint,revtex]{aps}
\begin{document}
\baselineskip=24pt plus 3pt minus 1pt
\hfill{\bf KFKI-RMKI-28-APR-1993}\bigskip\bigskip\bigskip\bigskip
\begin{title}
Unique Quantum Paths by Continuous Diagonalization
of the Density Operator
\end{title}
\author{Lajos Di\'osi}
\begin{instit}
KFKI Research Institute for Particle and Nuclear Physics\\
H-1525 Budapest 114, POB 49, Hungary\\
e-mail: diosi@rmki.kfki.hu
\end{instit}
\begin{abstract}
In this short note we show that for a Markovian open quantum system it is
always possible to construct a unique set of perfectly consistent
Schmidt paths, supporting quasi-classicality. Our Schmidt process,
elaborated several years ago, is the $\Delta t\rightarrow 0$ limit of
the Schmidt chain constructed very recently by Paz and Zurek.
\end{abstract}
\bigskip
\bigskip

\section{Introduction}

In a very recent work,
Paz and Zurek~[1] discuss Markovian open quantum
systems and construct Schmidt paths showing exact decoherence.
The authors,
however, notice their Schmidt paths are quite unstable under, e.g.,
varying the number of subsequent projections. They think to eliminate
the problem by tuning time intervals between subsequent projections
larger than the typical decoherence time.

In the present short note
I propose the opposite. Let time intervals be infinitely short!
It means that the frequency of the subsequent projections is so high that
the Schmidt path will be defined for any instant $t$ in the period
considered. This infinite frequency limit exists
and provides unique consistent set of Schmidt
paths. The issue was discussed~[2]
and completely solved~[3] several years ago.
Of course, the above limit is only valid up to Markovian approximation.
In fact, the "infinitesimal" repetition interval is still longer than
the response time of the reservoir.

For the sake of better distinction between ordinary~[1] and hereinafter
advocated~[3] Schmidt paths,
let us call them Schmidt {\it chain} and Schmidt {\it process}, respectively.

In the next Sect., Schmidt chains are briefly reviewed.The Sect. III.
will recall the earlier results available now for Schmidt processes.
Subsequently, in Sect. IV., we propose an application of Schmidt processes
to the quantum Brownian motion where classicality might be demonstrated.

\section{Schmidt Path---Markov Chain}

Consider the reduced dynamics of a given subsystem:
\begin{equation}
\rho(t^{\prime})=J(t^{\prime}-t)\rho(t),\ \ \ \ \ (t^{\prime}>t)
\end{equation}
where $\rho$ is the reduced density operator, $J$ is the Markovian
evolution superoperator. For a given sequence $t_0<t_1<\dots<t_n$
of selection times, let us have the corresponding sequence of pure
state (Hermitian) projectors: $\{P^0,P^1,\dots,P^n\}$ is a {\it Schmidt
chain} if
\begin{equation}
[P^{k+1},J(t_{k+1}-t_k)P^k]=0,\ \ \ \ \ k=0,1,2,\dots,n-1
\end{equation}
cf. Sect. 4 of Ref.~[1].
For fixed initial state $P^0=\rho(t_0)$, let the probabilities
\begin{equation}
p(P^1,P^2,\dots,P^n)=
tr\left[P^nJ(t_n-t_{n-1})P^{n-1}\dots P^1J(t_1-t_0)\rho(t_0)\right]
\end{equation}
be assigned to Schmidt chains. Schmidt chains satisfy the following
sum rule:
\begin{eqnarray}
\sum_{Schmidt\ paths}p(P^1,P^2,\dots,P^n)
      P^1\otimes P^2\otimes\dots\otimes P^n\nonumber\\
      =\rho(t_1)\otimes\rho(t_2)\otimes\dots\otimes\rho(t_n)
\end{eqnarray}
assuring the {\it consistency}~[4]
of the probability assignments (3).

Let us construct concrete Schmidt chains. To satisfy the Eq.\ (2) for $k=0$,
let us first diagonalize the positive definite operator $J(t_1-t_0)P^0$:
\begin{equation}
J(t_1-t_0)P^0=\sum_{\alpha}p^1_{\alpha}P^1_{\alpha}.
\end{equation}
If our choice is $P^1=P^1_{\alpha_1}$, whose probability is $p_{\alpha_1}$,
consider Eq.\ (2) for $k=1$ and diagonalize $J(t_2-t_1)P_{\alpha_1}^1$:
\begin{equation}
J(t_2-t_1)P^1_{\alpha_1}=\sum_{\alpha}p^2_{\alpha}P^2_{\alpha}.
\end{equation}
Single out $P^2=P_{\alpha_2}^2$ at random, with probability $p_{\alpha_2}^2$,
etc.

For the Schmidt chain
$\{P^0,P^1_{\alpha_1},P^2_{\alpha_2},\dots,P^n_{\alpha_n}\}$
one generates from the fixed initial state $P^0$,
the probability (3) takes the following factorized form:
\begin{equation}
p(P^1_{\alpha_1},P^2_{\alpha_2},\dots,P^n_{\alpha_n})\equiv
p(\alpha_1,\alpha_2,\dots,\alpha_n)
=p^1_{\alpha_1}p^2_{\alpha_2}\dots p^n_{\alpha_n}.
\end{equation}

Schmidt chain is Markov chain. Given $P^0=\rho(t_0)$, it will branch
at $t_1$ into $P^1_{\alpha_1}$,
i.e., into one of the eigenstate projectors of $J(t_1-t_0)P^0$;
the branching probability $p^1_{\alpha_1}$ is the corresponding eigenvalue.
In the general case, $P_{\alpha_k}^k$ will branch into
$P_{\alpha_{k+1}}^{k+1}$, i.e., into a certain eigenstate of
$J(t_{k+1}-t_k)P_{\alpha_k}^k$, with branching probability
$p_{\alpha_{k+1}}^{k+1}$ given by the corresponding eigenvalue.

\section{Schmidt path---Markov Process}

In this Sect., we consider the limiting case of the Schmidt chains when
the separations $t_{k+1}-t_k$ go to zero. The pure state path
$\{P(t);t>t_0\}$, starting from the fixed initial state
$P(t_0)=\rho(t_0)$, is a {\it Schmidt process}
if (for $t>t_0$ and $\epsilon\equiv dt>0$) $P(t+\epsilon)$ branches
into an eigenstate projector $P_\alpha(t)$ of
$J(\epsilon)P(t)$ while the branching
probability is the corresponding eigenvalue $p_\alpha(t)$.
Branching rates $w_\alpha(t)$ are worthwhile to introduce by
$p_\alpha(t)=\epsilon w_\alpha(t)$.

We follow the general results obtained in Ref.~[3]. Let us
introduce the Liouville superoperator $L$ generating the
Markovian evolution (1):
\begin{equation}
J(\epsilon)=1+\epsilon L.
\end{equation}
Assume the Lindblad form~[5]:
\begin{equation}
L\rho=-i[H,\rho]-{1\over2}\sum_\lambda
\left(F^\dagger_\lambda F_\lambda\rho+\rho F^\dagger_\lambda F_\lambda
      -2F_\lambda\rho F^\dagger_\lambda\right)
\end{equation}
where $H$ is the Hamiltonian and $\{F_\lambda\}$ are the Lindblad
generators. Following the method of Ref.~[3], introduce
the {\it frictional} (i.e. nonlinear-nonhermitian) {\it Hamiltonian}:
\begin{eqnarray}
H_P=H&-&{1\over2i}\sum_\lambda
\left(<F^\dagger_\lambda>F_\lambda-H.C.\right)\nonumber\\
     &-&{i\over2}\sum_\lambda
\left(F^\dagger_\lambda-<F^\dagger_\lambda>\right)
\left(F_\lambda-<F_\lambda>\right)+{i\over2}w
\end{eqnarray}
and the nonlinear positive definite {\it transition rate operator}:
\begin{equation}
W_P=\left(F_\lambda-<F_\lambda>\right)P
    \left(F^\dagger_\lambda-<F^\dagger_\lambda>\right)
\end{equation}
where, e.g., $<F_\lambda>\equiv tr(F_\lambda P)$.
We need the unit expansion of the transition rate operator:
\begin{equation}
W_P=\sum^\infty_{\alpha=1}w_{\alpha}P_\alpha.
\end{equation}
Observe that, due to the identity $W_PP\equiv0$, each $P_\alpha$ is
orthogonal to $P$.
The $w_\alpha$'s are called {\it transition (branching) rates}. The total
transition (branching) rate then follows from Eqs.\ (11) and (12):
\begin{equation}
w\equiv\sum_\alpha w_\alpha
=<F^\dagger_\lambda F_\lambda>-<F^\dagger_\lambda><F_\lambda>.
\end{equation}

How to generate Schmidt processes?
Given the initial pure state $\rho(t_0)=P(t_0)$, the pure state
$P(t)$ evolves according to the deterministic frictional
Schr\"odinger-von Neumann equation:
\begin{equation}
{d\over dt}P=-i(H_PP-PH_P^\dagger)
\end{equation}
except for discrete {\it orthogonal jumps} (branches)
\begin{equation}
P(t+0)=P_{\alpha}(t)
\end{equation}
occurring
from time to time at random with $P(t)$-dependent partial transition
rates $w_{\alpha}(t)$.  It is worthwhile to note that
neither $H_P$ nor $W_P$ depend on the concrete Lindblad representation (9)
of $L$, as it is clear in Ref.~[3].

Mathematically, the above Schmidt path
is  pure-state-valued Markov process of generalized Poissonian type.
During a given infinitesimal period $(t,t+dt)$,
the probability of the branch-free (i.e., jump-free, continuous)
evolution is $1-w(t)dt$. Consequently, one obtains~[6]
the {\it a priori} probability
of continuous evolution for an arbitrarily given period $(t_1,t_2)$ as
\begin{equation}
exp\left(-\int_{t_1}^{t_2}w(t)dt\right).
\end{equation}

\section{Classicality}

Schmidt processes assure maximum classicality in "measurement situations".
It has been shown in Ref.~[6] that for large enough t,
Schmidt process converges
to one of the pointer states while the overall probability of further
branches tends to zero. Convergence is then dominated by the asymptotic
solutions of the deterministic frictional Schr\"odinger-von Neumann Eq.\ (14).

To test classicality of Schmidt processes in less artificial situations,
let us start with the (modified~[7])
Caldeira-Leggett~[8] master equation:
\begin{eqnarray}
{d\over dt}\rho=L\rho=&-&i{1\over2M}[p^2,\rho]
                       -i\gamma[q,\{p,\rho\}]\nonumber\\
                      &-&{1\over2}\gamma\lambda_{dB}^{-2}[q,[q,\rho]]
                        -{1\over2}\kappa\gamma\lambda_{dB}^2[p,[p,\rho]]
\end{eqnarray}
where $\gamma$ is (two times) the friction constant, $\lambda_{dB}$ stands
for the thermal deBroglie length of the Brownian particle of mass M.
In Ref.~[7]
the value $\kappa=4/3$ has been suggested. For simplicity's, we have
omitted the usual renormalized potential term in the Hamiltonian,
assuming it is zero or small enough. Hence we can model quantum
counterpart of pure frictional motion.

Applying mechanically the Eqs.\ (10) and (11), we calculate both the frictional
Hamiltonian and the transition rate operator:
\begin{eqnarray}
H_P={1\over2M}p^2&+&{1\over2}\gamma\{q-<q>,p-<p>\}\nonumber\\
  &-&{i\over2}\gamma\left(\lambda_{dB}^{-2}((q-<q>)^2-\sigma_{qq}^2)+
                    \kappa\lambda_{dB}^2((p-<p>)^2-\sigma_{pp}^2)\right),
\end{eqnarray}
\begin{eqnarray}
W_{P}=\gamma\lambda_{db}^{-2}(q-<q>)P(q-<q>)
      &+&\kappa\gamma\lambda_{db}^2(p-<p>)P(p-<p>)\nonumber\\
      -i\gamma\Bigl((q-<q>)P(p-<p>)&-&(p-<p>)P(q-<q>)\Bigr)
\end{eqnarray}
where $\sigma_{qq}^2=<q^2>-<q>^2$ and $\sigma_{pp}^2=<p^2>-<p>^2$.
The total transition (branching) rate (13) obtains the simple form:
\begin{equation}
w=\gamma(\lambda_{dB}^{-2}\sigma_{qq}^2+\kappa\lambda_{dB}^2\sigma_{pp}^2-1)
\end{equation}
as can be easily verified by observing $w=trW_P$.

For most of the time the Schmidt process is governed by the frictional
Hamiltonian (18), via the nonlinear Eq.\ (14).
This equation {\it itself} possesses a stationary
solution $P(\infty)$ with simple Gaussian wave function
representing a standing particle.
Furthermore, one can heuristically guess that the nonhermitian terms
establish quasi-classicality for arbitrarily given initial states.
Obviously, the random jumps (15) will interrupt the deterministic
evolution of the Schmidt process. If, however, in the quasi-classical regime
the rate (20) of jumps were much smaller than the speed of
relaxation due to the frictional Hamiltonian (18) then the infrequent
jumps would only cause slight random walk and breathing
to the otherwise quasi-classical wave function.

\section{Conclusion}

Schmidt processes offer a certain solution to the preferred basis problem of
quantum mechanics, at least when the subsystem's reduced dynamics can be
considered Markovian. It will be interesting to carry on with analytic
calculations for the Schmidt process of the Brownian motion, not at all
exhausted in Sec. IV.

\bigskip
This work was supported by the Hungarian Scientific Research
Fund under Grant OTKA 1822/1991.

\end{document}